\documentclass{PoS}
\usepackage{amsmath}
\usepackage{bm}
\usepackage{cite}
\title{
Equation of state for hybrid stars with strangeness
}
\ShortTitle{
Equation of state for hybrid stars with strangeness
}
\author{
\speaker{Tsuyoshi Miyatsu}, Takahide Kambe, and Koichi Saito\\
Department of Physics, Faculty of Science and Technology, Tokyo University of Science, Noda 278-8510, Japan\\
E-mail: \email{tsuyoshi.miyatsu@rs.tus.ac.jp},
\email{koichi.saito@rs.tus.ac.jp}
}
\abstract{
Considering the mass constraint from the resent pulsar observations, we study the properties of neutron stars including hyperons and quarks explicitly.
Using the chiral quark-meson coupling model with relativistic Hartree-Fock approximation, the equation of state (EoS) for hadronic matter is calculated by taking into account the strange ($\sigma^{\ast}$ and $\phi$) mesons as well as the light non-strange ($\sigma$, $\omega$, $\bm{\rho}$, and $\bm{\pi}$) mesons in SU(3) flavor symmetry. On the other hand, the EoS for quark matter is constructed with the simple MIT bag or the flavor-SU(3) Nambu-Jona-Lasinio model, and we investigate the effect of the hadron-quark coexistence on the neutron-star properties, imposing smooth crossover or Gibbs criterion for chemical equilibrium.
The mass-radius relation of a neutron star, as well as physical quantities such as EoSs, particle fractions, and the speed of sound in matter are presented.
We find that, in order to prevent the quark appearance at very low densities, the stiff hadronic EoS should be required under both of the hadron-quark crossover and the first-order phase transition.
}
\FullConference{
The 26th International Nuclear Physics Conference\\
11-16 September, 2016\\
Adelaide, Australia
}
\begin{document}

\section{Introduction}
\label{sec:introduction}

Neutron stars may be believed to be cosmological laboratories for nuclear matter at extremely low temperature and high density, because the central density of neutron stars can reach several times higher than the normal nuclear density.
The neutron-star properties, especially the mass and radius, are particularly important for determining the equations of state (EoSs) in the core of a neutron star.
Thus, the astrophysical observations of pulsars can provide some constraints on the EoS for dense nuclear matter~\cite{Weber:1999qn,Glendenning:2000,Lattimer:2006xb}.
The possibility of the other exotic degrees of freedom are also expected in the core of a neutron star, such as hyperons~\cite{Glendenning:1991es,Schaffner:1995th}, quark matter~\cite{Kettner:1994zs,Weber:2004kj}, meson condensations~\cite{Takatsuka:1978ku,Glendenning:1998zx}, and/or dark matter~\cite{PerezGarcia:2010ap}.

Since the precise observations of massive neutron stars with the mass around $2M_{\odot}$~\cite{Demorest:2010bx,Antoniadis:2013pzd},
it is quite difficult to explain them theoretically,
because the new degrees of freedom, especially hyperons, soften the EoS for neutron stars, and thus, the possible maximum mass of a neutron star is drastically reduced.
Hence, it seems to be important to solve the problem by proposing the neutron-star EoS which can satisfy the nuclear properties and the astrophysical constraints.

In the present study, the EoS for neutron stars with strangeness is presented.
We first show the hadronic EoS with hyperons, which can support $2M_{\odot}$ neutron stars, using the relativistic mean-field (Hartree) or Hartree-Fock (RHF) approximation in SU(3) flavor symmetry~\cite{Miyatsu:2013yta,Miyatsu:2014wca,Weissenborn:2011ut,Lopes:2013cpa,Miyatsu:2011bc,Katayama:2012ge,Miyatsu:2013hea,Whittenbury:2013wma}.
The EoS for quark matter is then constructed with the simple MIT bag or the flavor-SU(3) Nambu--Jona-Lasinio (NJL) model, and we investigate the effect of the transition between hadron and quark phases on the neutron-star properties, imposing smooth crossover or Gibbs criterion for chemical equilibrium~\cite{Miyatsu:2015kwa,Kambe:2016olv,Whittenbury:2015ziz}.

\section{Hadronic EoS for neutron stars with hyperons}
\label{sec:hadronic-EoS}

Lagrangian density for uniform hadronic matter is given by
\begin{equation}
  \label{eq:Lagrangian}
  \mathcal{L}_{H} = \mathcal{L}_{B} + \mathcal{L}_{M} + \mathcal{L}_{\rm int},
\end{equation}
%
with the baryon, meson, and interaction terms~\cite{Miyatsu:2011bc,Katayama:2012ge}.
We here consider the octet baryons ($B$): proton ($p$), neutron ($n$), $\Lambda$, $\Sigma^{+0-}$, and $\Xi^{0-}$.
In addition, the mesons, which are composed of light quarks ($\sigma$, $\omega$, $\bm{\rho}$, and $\bm{\pi}$) and the strange quarks ($\sigma^{\ast}$ and $\phi$), are taken into account.
The interaction Lagrangian density is given by
\begin{align}
  \label{eq:interaction-Lagrangian}
  \mathcal{L}_{\rm int}
  & = \sum_{B}\bar{\psi}_{B} \biggl[ g_{\sigma B}\left(\sigma\right)\sigma
  +   g_{\sigma^{\ast}B}\left(\sigma^{\ast}\right)\sigma^{\ast}
  -   g_{\omega B}\gamma_{\mu}\omega^{\mu}
  +   \frac{f_{\omega B}}{2\mathcal{M}}\sigma_{\mu\nu}\partial^{\nu}\omega^{\mu}
  -   g_{\phi B}\gamma_{\mu}\phi^{\mu}
  \nonumber \\
  & + \frac{f_{\phi B}}{2\mathcal{M}}\sigma_{\mu\nu}\partial^{\nu}\phi^{\mu}
  -   g_{\rho B}\gamma_{\mu}\bm{\rho}^{\mu}\cdot\bm{I}_{B}
  +   \frac{f_{\rho B}}{2\mathcal{M}}\sigma_{\mu\nu}\partial^{\nu}\bm{\rho}^{\mu}\cdot\bm{I}_{B}
  -   \frac{f_{\pi B}}{m}\gamma_{5}\gamma_{\mu}\partial^{\mu}\bm{\pi}\cdot\bm{I}_B \biggr] \psi_{B},
\end{align}
where the common mass scale mass, $\mathcal{M}$, is taken to be the free nucleon mass, and $\bm{I}_B$ is the isospin matrix for baryon $B$.
The $\sigma$-, $\sigma^{\ast}$-, $\omega$-, $\phi$-, $\rho$-, $\pi$-$B$ coupling constants are respectively denoted by $g_{\sigma B}(\sigma)$, $g_{\sigma^{\ast}B}(\sigma^{\ast})$, $g_{\omega B}$, $g_{\phi B}$, $g_{\rho B}$, and $f_{\pi B}$, while $f_{\omega B}$, $f_{\phi B}$, and $f_{\rho B}$ are the tensor coupling constants for the vector mesons.

In order to consider the baryon structure variation in nuclear matter, we adopt the following simple parametrizations for the $\sigma$- and $\sigma^{\ast}$-$B$ coupling constants, which are based on the chiral quark-meson coupling (chiral QMC, CQMC) model~\cite{Nagai:2008ai,Miyatsu:2010zz,Saito:2010zw}:
\begin{equation}
  \label{eq:cc-scalar}
  g_{\sigma B}(\sigma)
  = g_{\sigma B}b_{B}\left[1-\frac{a_{B}}{2}\left(g_{\sigma N}\sigma\right)\right],
  \hspace{0.02\textwidth}
  g_{\sigma^{\ast}B}(\sigma^{\ast})
  = g_{\sigma^{\ast}B}b_{B}^{\prime}\left[1-\frac{a_{B}^{\prime}}{2}\left(g_{\sigma^{\ast}\Lambda}\sigma^{\ast}\right)\right],
\end{equation}
where $a_{B}$, $a_{B}^{\prime}$, $b_{B}$, and $b_{B}^{\prime}$ are parameters listed in Refs~\cite{Miyatsu:2013yta,Miyatsu:2013hea,Miyatsu:2015kwa,Miyatsu:2014wca}.
In contrast, the nonlinear (NL) self-interaction terms for the $\sigma$ field are introduced in quantum hadrodynamics (QHD)~\cite{Boguta:1977xi}, where the baryons are treated as point-like objects, and the coupling constants for the scalar mesons are fixed as $g_{\sigma B}(\sigma=0)$ and $g_{\sigma^{\ast}B}(\sigma^{\ast}=0)$ in Eq.~\eqref{eq:interaction-Lagrangian}.

The nucleon coupling constants are determined so as to reproduce the same saturation condition: the binding energy per nucleon ($-16.1$ MeV) and the symmetry energy ($32.5$ MeV) at the normal nuclear density, $\rho_{0}=0.155$ fm$^{-3}$.
In addition, the coupling constants for hyperons are also fixed to simulate the experimental data of hypernuclei in SU(3) flavor symmetry or SU(6) spin-flavor symmetry~\cite{Miyatsu:2013yta,Miyatsu:2014wca,Weissenborn:2011ut,Lopes:2013cpa}.

\begin{figure}[t!]
\centering
\includegraphics[width=.60\textwidth]{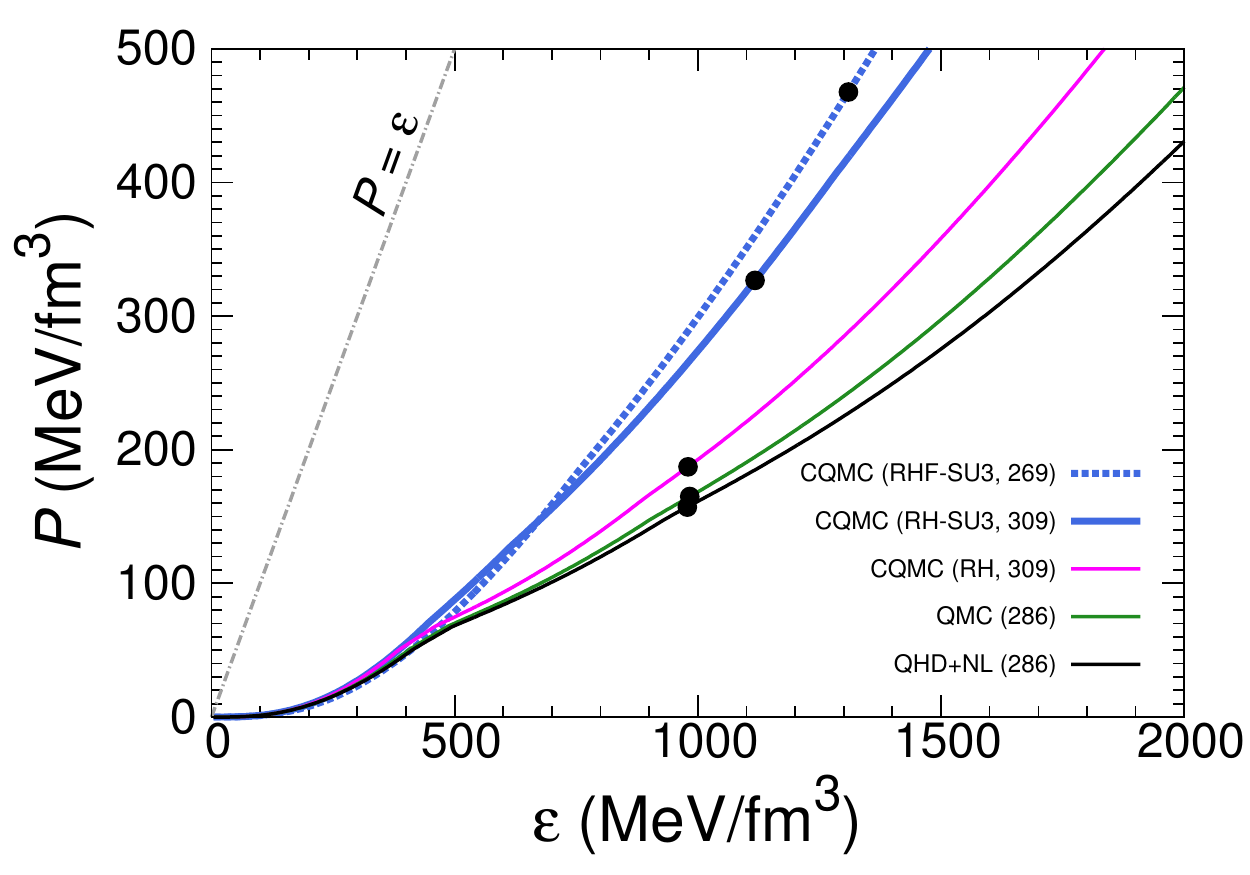}
\includegraphics[width=.36\textwidth]{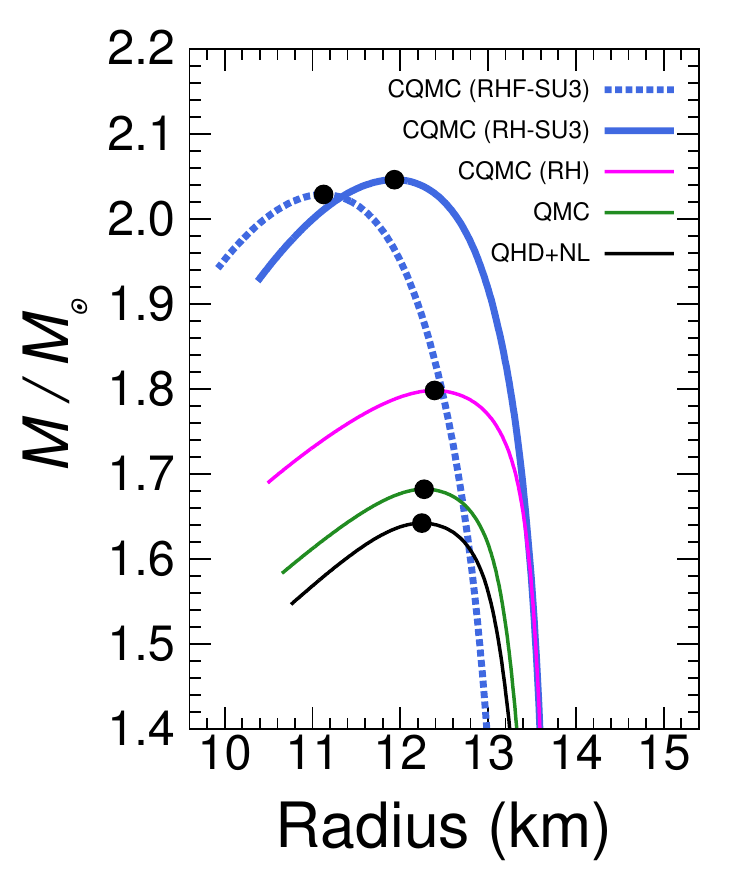}
\caption{EoS for neutron stars with hyperons (left panel) and mass-radius relation (right panel).
The label denoted by ``SU3'' is for the case in SU(3) flavor symmetry with relativistic Hartree (RH) or Hartree-Fock (RHF) approximation, while the lines without the label are calculated in SU(6) spin-flavor symmetry with RH approximation.
The filled circle shows the point at which a neutron star reaches the maximum mass.
The incompressibility of symmetric nuclear matter at $\rho_{0}$ is also shown by the last number in each label (left panel).}
\label{fig:hadron}
\end{figure}
In Fig.~\ref{fig:hadron}, we show the EoS for neutron stars with hyperons and the mass-radius relation of a neutron star by solving the Tolman-Oppenheimer-Volkoff (TOV) equation, where leptons ($\ell$) are introduced to impose the charge neutrality and $\beta$ equilibrium conditions.
We find that, due to the variation of the quark substructure of baryon in matter, the EoS in the QMC model is slightly stiffer than that in the QHD+NL model at the high energy densities.
In addition, the effect of gluon and pion exchanges between quarks in the CQMC model makes the EoS harder.
However, it is impossible to support $2M_{\odot}$ neutron stars even using the CQMC model in SU(6) spin-flavor symmetry.
With relativistic Hartree (RH) or Hartree-Fock (RHF) approximation in SU(3) flavor symmetry, the maximum mass can exceed the astrophysical mass constraint, because the additional repulsive force due to the $\phi$ meson makes the EoS stiff, even if hyperons are taken into account in the core.
We note that, though the EoSs in SU(3) symmetry can satisfy the $2M_{\odot}$ constraint from the astrophysical observations, the incompressibility of symmetric nuclear matter at $\rho_{0}$ with the RHF approximation shows a more reasonable value, $K_{0}=269$ MeV, than that with the RH approximation, $K_{0}=309$ MeV.
Furthermore, we find that the Fock contribution reduces the radius of a neutron star maximally by about 1 km.

\section{Quark matter description and hadron-quark coexistence}
\label{sec:phase-transition}

For a description of quark matter, we use two models: one is the simple MIT bag model with the density-dependent bag constant, which is assumed to be given by a Gaussian parametrization,
\begin{equation}
  \label{eq:DDbagcc}
  B(\rho) = B_{\infty} + \left(B_{0}-B_{\infty}\right)\exp\left[-\beta\left(\frac{\rho}{\rho_{0}}\right)^{2}\right],
\end{equation}
with $\rho$ and $\beta$ being the total baryon density and a adjustable parameter, respectively~\cite{Miyatsu:2015kwa}.
The other is the flavor-SU(3) Nambu--Jona-Lasinio (NJL) model,
\begin{align}
  \label{eq:NJL-Lagrangian}
  \mathcal{L}_{\rm NJL}
  & = \bar{q}\left(i\gamma_{\mu}\partial^{\mu}-\widehat{m}\right)q
  +   \frac{1}{2}G_{S}\sum_{a=0}^{8}\left[\left(\bar{q}\lambda^{a}q\right)^{2}
  +   \left(\bar{q}i\gamma_{5}\lambda^{a}q\right)^{2}\right] \nonumber \\
  & - G_{D}\left[\rm{det}_{f}\left(\bar{q}(1+\gamma_{5})q\right)
  +   \rm{det}_{f}\left(\bar{q}(1-\gamma_{5})q\right)\right]-\frac{1}{2}g_{V}\left(\bar{q}\gamma^{\mu}q\right)^{2},
\end{align}
where the quark field, $q_{i}$ ($i=u,d,s$), has three colors and three flavors with the current quark mass, $m_{i}$, in the mass matrix, $\widehat{m}$.
We here introduce a phenomenological vector-type interaction, and the parameter set given in Ref.~\cite{Hatsuda:1994pi} is used.

In order to describe the hadron-quark mixture in the core, we consider the first-order phase transition under $\beta$-equilibrium with the so-called Gibbs criterion for chemical equilibrium~\cite{Miyatsu:2015kwa}, or the crossover phenomenon~\cite{Kambe:2016olv}.
In addition, we use the following three cases for the hadron-quark crossover~\cite{Masuda:2012kf,Masuda:2012ed,Hell:2014xva}:
the energy density-baryon density ($\epsilon$-$\rho$) interpolation,
the pressure-baryon density ($P$-$\rho$) interpolation,
and the pressure-energy density ($P$-$\epsilon$) interpolation.
In all cases, we employ the common interpolation functions:
\begin{equation}
  \label{eq:interpolation-function}
  f_{\pm}(\rho)=\frac{1}{2}\left[1\pm\tanh\left(\frac{\rho-\bar{\rho}}{\Gamma}\right)\right]
  \hspace{0.02\textwidth}\text{or}\hspace{0.02\textwidth}
  f_{\pm}(\epsilon)=\frac{1}{2}\left[1\pm\tanh\left(\frac{\epsilon-\bar{\epsilon}}{\Gamma}\right)\right],
\end{equation}
%
with $\bar{\rho}$ ($\bar{\epsilon}$) and $\Gamma$ being the central density (energy density) and width for the crossover region, respectively.

\section{Hybrid-star properties}
\label{sec:hybrid-star}

\begin{figure}[t!]
\centering
\includegraphics[width=.48\textwidth]{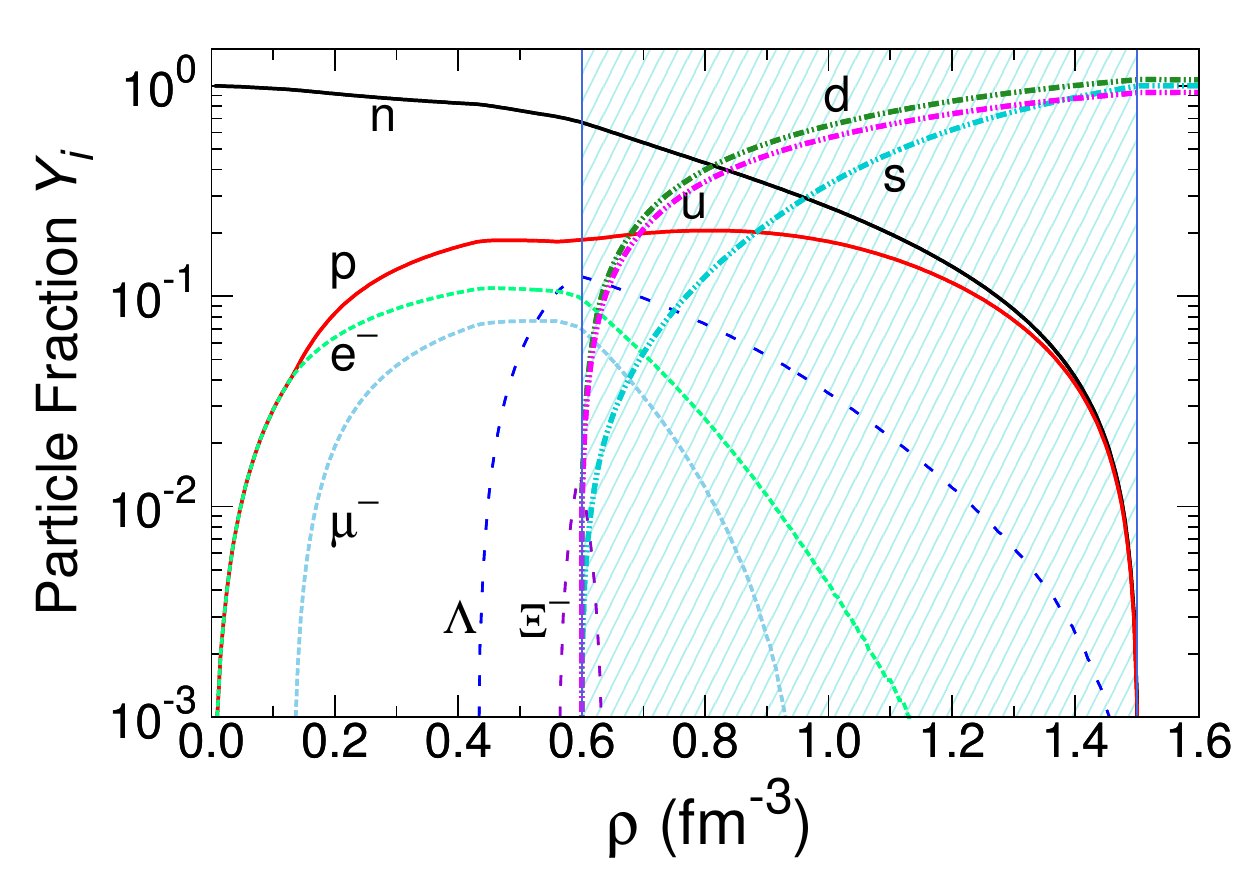}
\includegraphics[width=.48\textwidth]{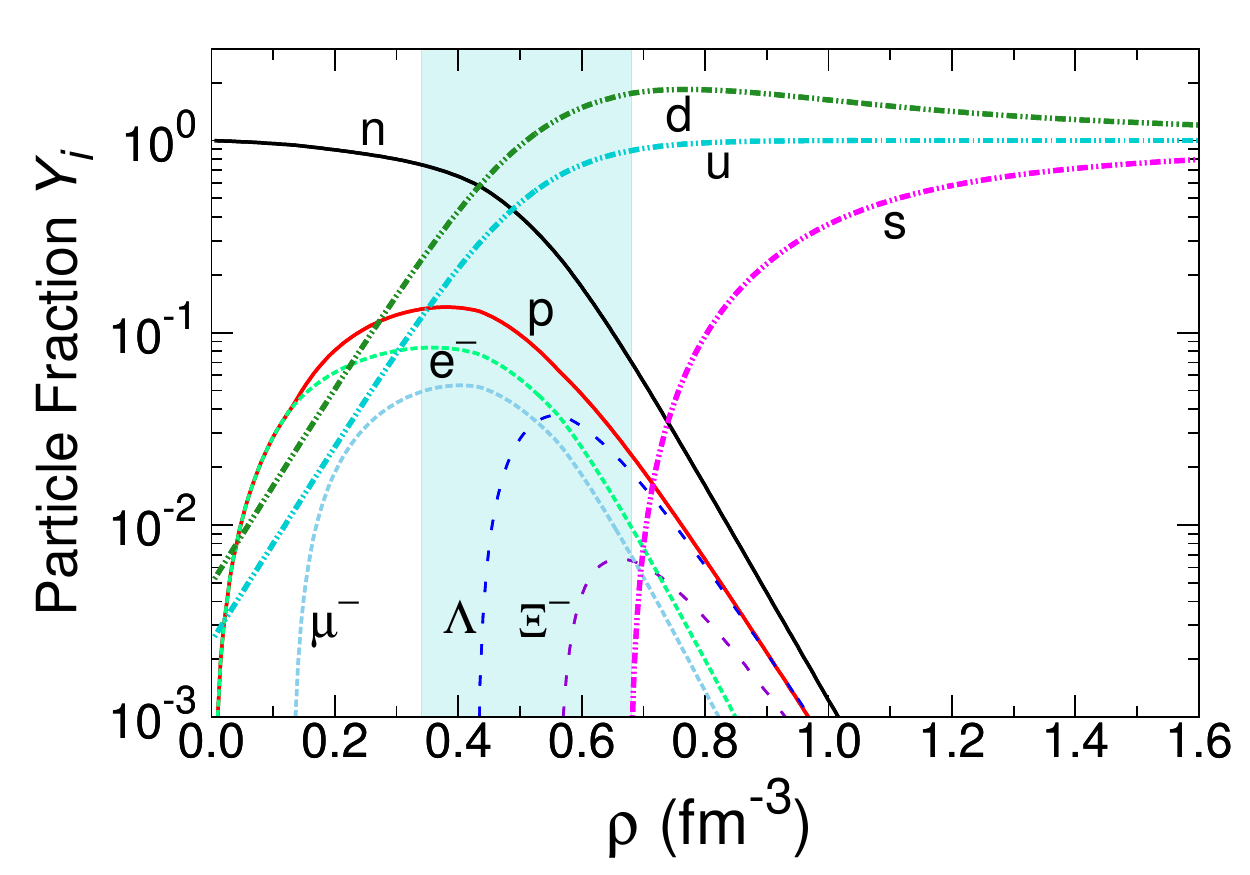}
\caption{Particle fraction, $Y_{i}$, for hybrid-star matter.
The left panel is for the case of the first-order phase transition with Gibbs criterion ($\beta=0.025$), and the right panel is for the case under crossover phenomenon in the $P$-$\rho$ interpolation with $\bar{\rho}=3\rho_{0}^{\rm NJL}$ and $\Gamma=\rho_{0}^{\rm NJL}$.
The hadronic EoS in the CQMC model with RH-SU3, as in Fig.~\ref{fig:hadron}, is used in both cases.
For the quark EoS, we employ the MIT bag (NJL) model in the left (right) panel.
The hatched area presents the mixed phase of hadrons and quarks, and the shaded area is the crossover region.}
\label{fig:Composition}
\end{figure}
\begin{figure}[t!]
\centering
\includegraphics[width=.524\textwidth]{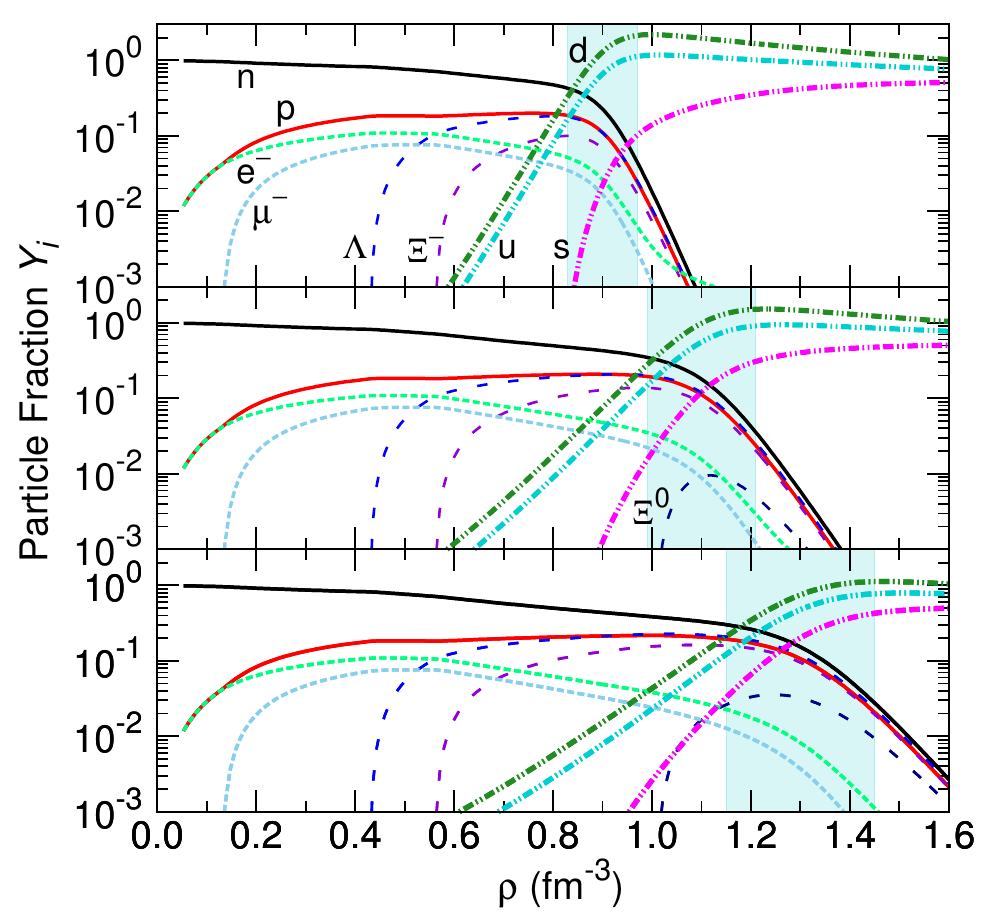}
\includegraphics[width=.470\textwidth]{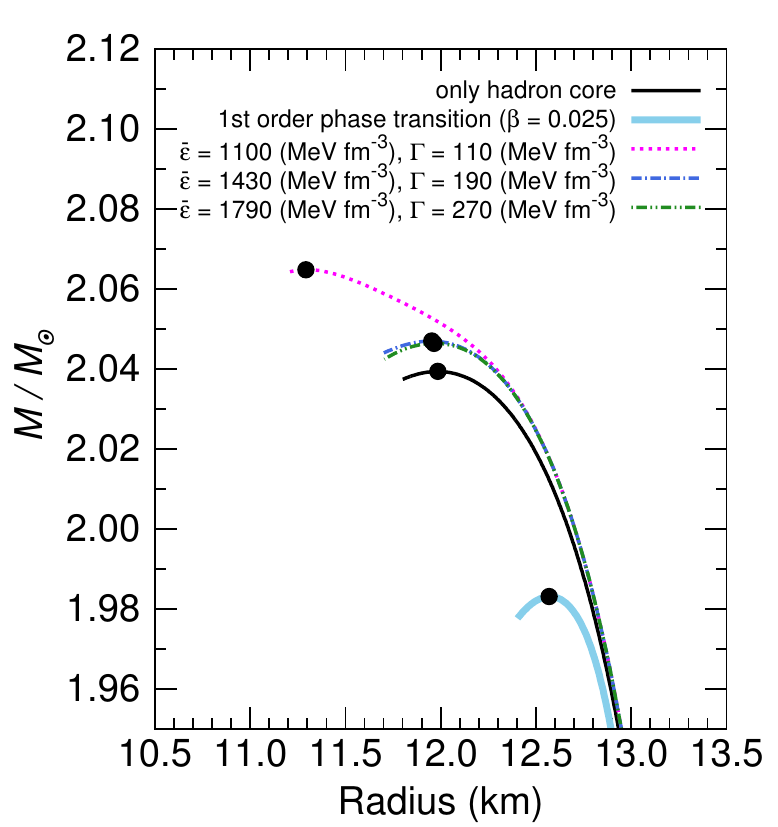}
\caption{Particle fraction, $Y_{i}$, for hybrid-star matter under crossover phenomenon in the $P$-$\epsilon$ interpolation (left panel), and mass-radius relation for a hybrid star (right panel).
We here adopt the CQMC model with RH-SU3 for the hadronic EoS and the flavor-SU(3) NJL model with $g_{V}=1.5G_{S}$ for the quark EoS in both panels.
The top (middle) [bottom] figure in the left panel is for the case with $\bar{\epsilon}=1100$ ($1430$) [$1790$] MeV/fm$^{3}$ and $\Gamma=110$ ($190$) [$270$] MeV/fm$^{3}$.}
\label{fig:press-engy-crossover}
\end{figure}
The particle fraction, $Y_{i}=\rho_{i}/\rho$ ($i=B,\ell,q$), for hybrid-star matter is presented in Fig.~\ref{fig:Composition}.
In case of the first-order phase transition shown in the left panel, the quarks appear at 0.6 fm$^{-3}$, following the hyperon creation.
Considering the $2M_{\odot}$ constraint from the astrophysical observations, the parameter in the density-dependent bag constant in Eq.~\eqref{eq:DDbagcc} should be $\beta\le 0.025$.
In addition, we require the stiff hadronic EoS using the MIT bag model for quark matter, because the quark appearance also softens the EoS for hybrid-star matter as well as the hyperon creation does.
In contrast, in the right panel (NJL model) of Fig.~\ref{fig:Composition}, the light quarks are seen even at very low densities, using the typical range for the crossover phenomenon in the $P$-$\rho$ interpolation, namely $\bar{\rho}=3\rho_{0}^{\rm NJL}$ and $\Gamma=\rho_{0}^{\rm NJL}$ ($\rho_{0}^{\rm NJL}=0.17$ fm$^{-3}$)~\cite{Masuda:2012kf,Masuda:2012ed}.
Here the number density of particle species, $\rho_{i}$, is defined as $\rho_{i}=f_{-}(\rho)\rho_{i}^{\rm HM}+f_{+}(\rho)\rho_{i}^{\rm QM}$ with $\rho_{i}^{\rm HM}$ ($\rho_{i}^{\rm QM}$) being that in hadronic (quark) matter.
Even if we use the soft hadronic EoS, the $2M_{\odot}$ constraint can be easily satisfied by using the NJL model for quark matter, because of the strong repulsive force due to the phenomenological vector-type interaction.

We next study how the crossover region affects the hybrid-star properties.
In order to prevent the quark creation at low densities, the crossover window should be moved to higher densities.
In the left panel of Fig.~\ref{fig:press-engy-crossover}, the particle fraction for hybrid-star matter in the $P$-$\epsilon$ interpolation with $g_{V}=1.5G_{S}$ is presented.
The central energy density and width for the crossover window are here determined
so as to achieve the appearance of quarks at $0.6$ fm$^{-3}$.
As the crossover region goes to high densities, the amount of strangeness increases and the $\Xi^{0}$ appears at around 1.0 fm$^{-3}$ in the middle and bottom figures of the left panel.
It means that the hadronic EoS dominates, and then the EoS for hybrid-star matter becomes soft.
Thus, the possible maximum mass of a hybrid star is slightly reduced as shown in the right panel of Fig.~\ref{fig:press-engy-crossover}.
We note that, in the case with $g_{V}\le G_{S}$, the maximum mass of a hybrid star is smaller than that of a neutron star without quarks.

\begin{figure}[t!]
\centering
\includegraphics[width=.60\textwidth]{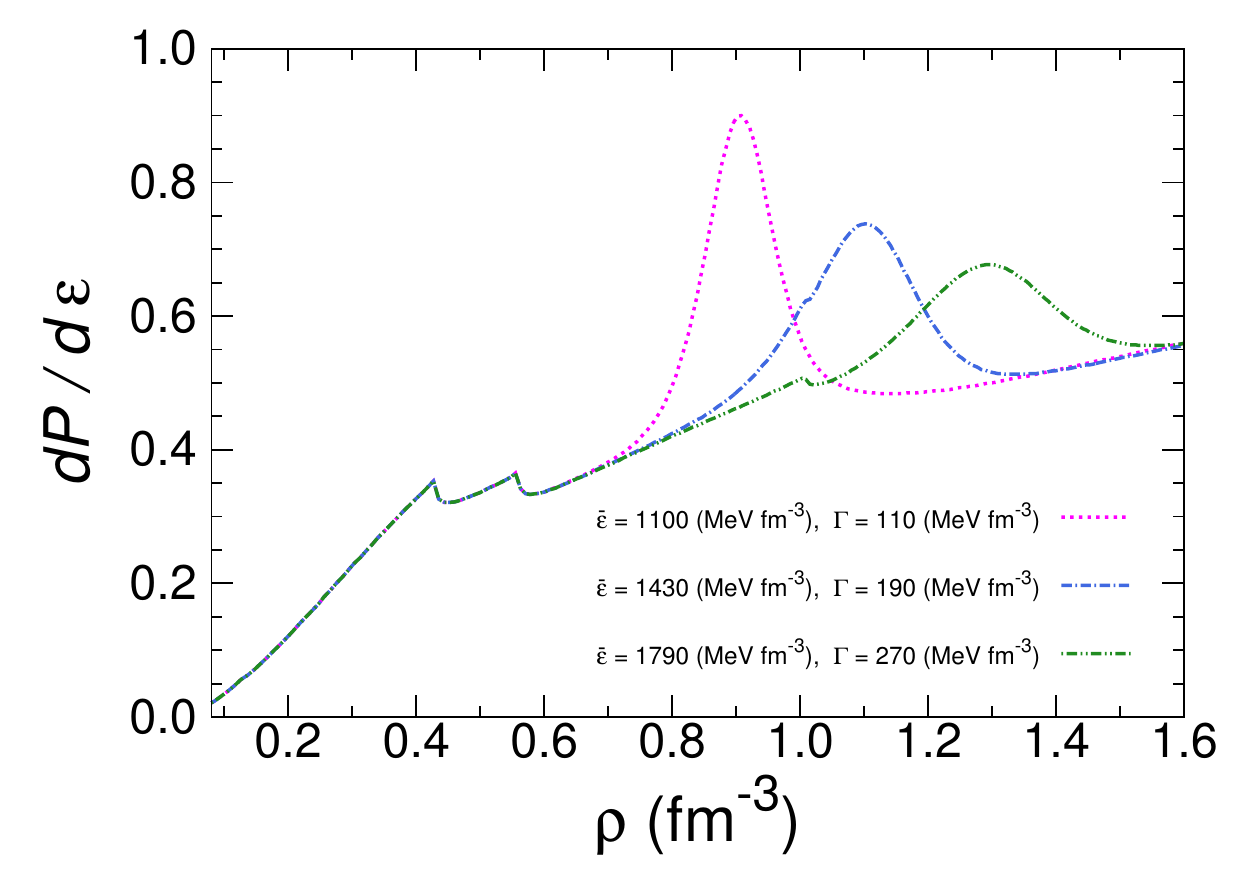}
\caption{Speed of sound as a function of baryon density in the $P$-$\epsilon$ interpolation with $g_{V}=1.5G_{S}$.
}
\label{fig:sound-velocity}
\end{figure}
The speed of sound in the $P$-$\epsilon$ interpolation with $g_{V}=1.5G_{S}$ is given in Fig.~\ref{fig:sound-velocity}.
In this case, causality and thermodynamics stability are fully  satisfied in any crossover window, namely $0<dP/d\epsilon<1$.
In addition, we find that these conditions can be satisfied in the $P$-$\epsilon$ and $P$-$\rho$ interpolations, if the crossover region is located at high density.
However, in the $P$-$\rho$ interpolation, it is impossible to explain the $2M_{\odot}$ constraint form the astrophysical observations.

\section{Summary}
\label{sec:summary}

We have calculated the equation of state (EoS) for neutron stars with hyperons and quarks explicitly, where the nuclear properties and the astrophysical constraints are considered.
The hadronic EoS for neutron-star matter has been calculated using the chiral quark-meson coupling (CQMC) model in order to include the effect of internal structure variation of baryons in matter.
In addition, not only the light non-strange ($\sigma$, $\omega$, $\bm{\rho}$, and $\bm{\pi}$) mesons but also the strange ($\sigma^{\ast}$ and $\phi$) mesons have been taken into account with relativistic Hartree-Fock (RHF) approximation.
The coupling constants for baryons are also determined so as to reproduce the saturation properties and the experimental data of hypernuclei in SU(3) flavor symmetry or SU(6) spin-flavor symmetry.
We have found that the baryon structure variation in matter, the Fock contribution, and the repulsive force due to the $\phi$ meson stiffen the EoS for neutron stars, and thus, the possible maximum mass of a neutron star can reach the $2M_{\odot}$ constraint from the astrophysical observations, even if we consider the hyperons in the core.

On the other hand, the EoS for quark matter has been constructed with the simple MIT bag model with the density-dependent bag constant or the flavor-SU(3) Nambu--Jona-Lasinio (NJL) model with phenomenological vector-type interaction.
The effect of hadron-quark coexistence on the hybrid-star properties have been investigated by imposing the smooth crossover and the first-order phase transition for chemical equilibrium.
It has been found that, using the MIT bag model with Gibbs criterion, the possible maximum mass of a hybrid star can reach $2M_{\odot}$, if we employ the stiff hadronic EoS.
Furthermore, we have also found that, using the hadron-quark crossover with the three kinds of interpolations in the NJL model, both of the astrophysical constraints and the speed of sound in hybrid-star matter can be satisfied only in the pressure-energy density ($P$-$\epsilon$) interpolation, if the crossover window is pushed upwards in order to prevent the creation of much amount of quarks at very low densities.

\end{document}